\documentclass[aps,prl,preprint,groupedaddress]{revtex4-1}

\usepackage{float}
\usepackage{hyperref}
\usepackage{epsfig}
\usepackage{graphicx}
\usepackage{subfigure}
\usepackage{latexsym}
\usepackage{color}
\usepackage{fullpage}
\usepackage{epstopdf}
\usepackage{dcolumn}
\usepackage{bm}
\usepackage{ulem}
\usepackage{units}
\usepackage{amsmath}
\usepackage{lineno}

\begin{document}

\title{Extensive degeneracy, Coulomb phase and magnetic monopoles in an artificial realization of the square ice model}

\author{Yann Perrin$^{1,2}$, Benjamin Canals$^{1,2}$ and Nicolas Rougemaille$^{1,2}$}

\affiliation{$^1$ CNRS, Inst NEEL, F-38000 Grenoble, France\\$^2$ Univ. Grenoble Alpes, Inst NEEL, F-38000 Grenoble, France}

\date{\today}

\begin{abstract}

\textbf{Artificial spin ice systems have been introduced as a possible mean to investigate frustration effects in a well-controlled manner by fabricating lithographically-patterned two-dimensional arrangements of interacting magnetic nanostructures. This approach offers the opportunity to visualize unconventional states of matter, directly in real space, and triggered a wealth of studies at the frontier between nanomagnetism, statistical thermodynamics and condensed matter physics. Despite the strong efforts made these last ten years to provide an artificial realization of the celebrated square ice model, no simple geometry based on arrays of nanomagnets succeeded to capture the macroscopically degenerate ground state manifold of the corresponding model. Instead, in all works reported so far, square lattices of nanomagnets are characterized by a magnetically ordered ground state consisting of local flux-closure configurations with alternating chirality. Here, we show experimentally and theoretically, that all the characteristics of the square ice model can be observed if the artificial square lattice is properly designed. The spin configurations we image after demagnetizing our arrays reveal unambiguous signatures of an algebraic spin liquid state characterized by the presence of pinch points in the associated magnetic structure factor. Local excitations, i.e. classical analogues of magnetic monopoles, are found to be free to evolve in a massively degenerated, divergence-free vacuum. We thus provide the first lab-on-chip platform allowing the investigation of collective phenomena, including Coulomb phases and ice-like physics.}

\end{abstract}

\maketitle

Artificial spin ices made from arrays of interacting magnetic nano-islands have received considerable attention over the last decade \cite{Nisoli2013, Heyderman2013, Cumings2014}. To a large extend, this attention is due to the capability of these systems to mimic the exotic many body physics of highly frustrated magnetic compounds and to visualize this physics directly, in real space. In particular, artificial spin ices have proven to be powerful simulators of the emergent phenomena that have so characterized frustrated magnetism. The imaging of monopole-like excitations \cite{Ladak2010, Mengotti2010} and charge crystallization \cite{Rougemaille2011, Zhang2013, Montaigne2014, Chioar2014-1, Drisko2015, Anghinolfi2015} in kagome lattices, the discovery of new magnetic textures \cite{Zhang2012, Chioar2014-2, Chioar2016} and the observation of magnetic moment fragmentation \cite{Brooks2014, Canals2016} are striking examples illustrating the strength of being capable to fabricate artificial magnetic architectures with desired specifications.

\begin{figure}[H]
\center
\includegraphics[width=12cm]{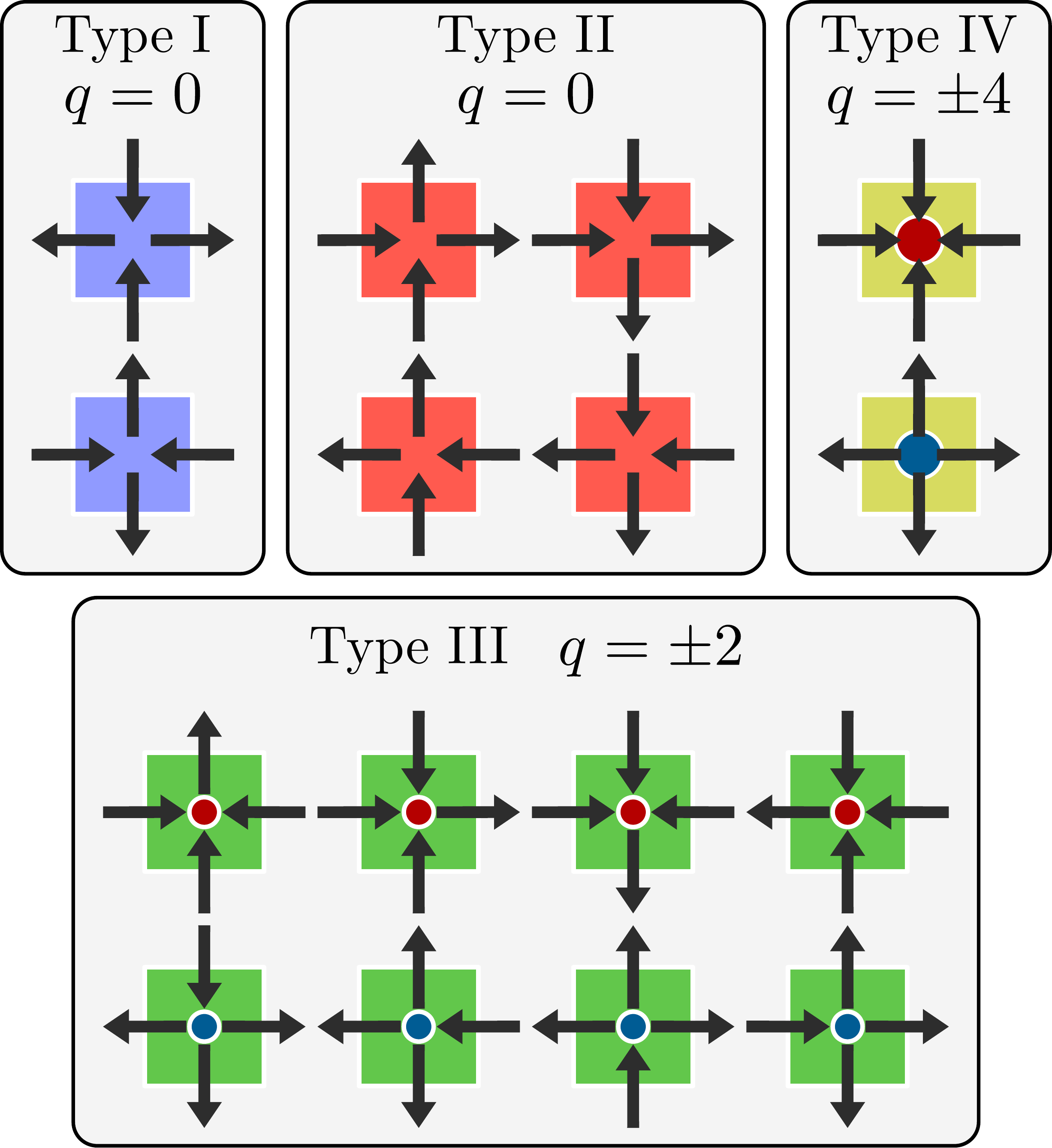}
\caption{\label{vertices} \textbf{Different vertex types of the six-vertex model.} Each vertex of the square lattice hosts one of the $2^4=16$ possible spin (black arrows) states. These 16 states can be classified in 4 vertex types, represented by squares colored in blue, red, green and yellow for type-I, II, III and IV vertices, respectively. Type-I (blue) and type-II (red) vertices correspond to a state in which two spins are pointing inwards and 2 spins are pointing outwards of the vertex. They are thus characterized by the absence of a magnetic charge $q$. Type-III (green) and type-IV (yellow) vertices correspond to spin states associated to a non zero magnetic charge ($q=\pm 2$ or $\pm 4$, respectively) represented by blue/red circles.}
\end{figure}

One of the very first motivations of fabricating an artificial magnetic ice was to mimic the intriguing properties of pyrochlore spin ice materials and to explore the physics of the associated six-vertex model \cite{Wang2006}. In two dimensions, this model is defined by a square lattice of (multiaxes) Ising pseudo-spins coupled through nearest-neighbor interactions only. Each vertex of the lattice has then a coordination number of 4 and hosts one of the $2^4=16$ possible spin states. These 16 configurations can be classified in 4 different types [see Fig. \ref{vertices}], having possibly 4 different energies. Among these 16 states, 6 obey the so-called 'ice rule' consisting in two spins pointing in and two spins pointing out at each vertex of the lattice (type-I and type-II vertices). In the particular case where type-I and type-II vertices have the same minimum energy, while type-III and type-IV vertices correspond to high-energy states, this six-vertex model is known as the square ice model \cite{Lieb1967}.

The beauty of the square ice model, initially introduced to describe ice physics, is to capture most of the astonishing properties of rare earth compounds, such as dysprosium and holmium titanates \cite{Fennell2009}: the low-energy manifold is macroscopically degenerated, meaning that the system still highly fluctuates at very low temperature \cite{Lieb1967}. This liquid-like behavior between strongly constrained, divergence-free spin states leads to a Coulomb phase \cite{Henley2010} and magnetic excitations involving fractional quasi-particles \cite{Castelnovo2008}.

\begin{figure}[H]
\center
\includegraphics[width=13cm]{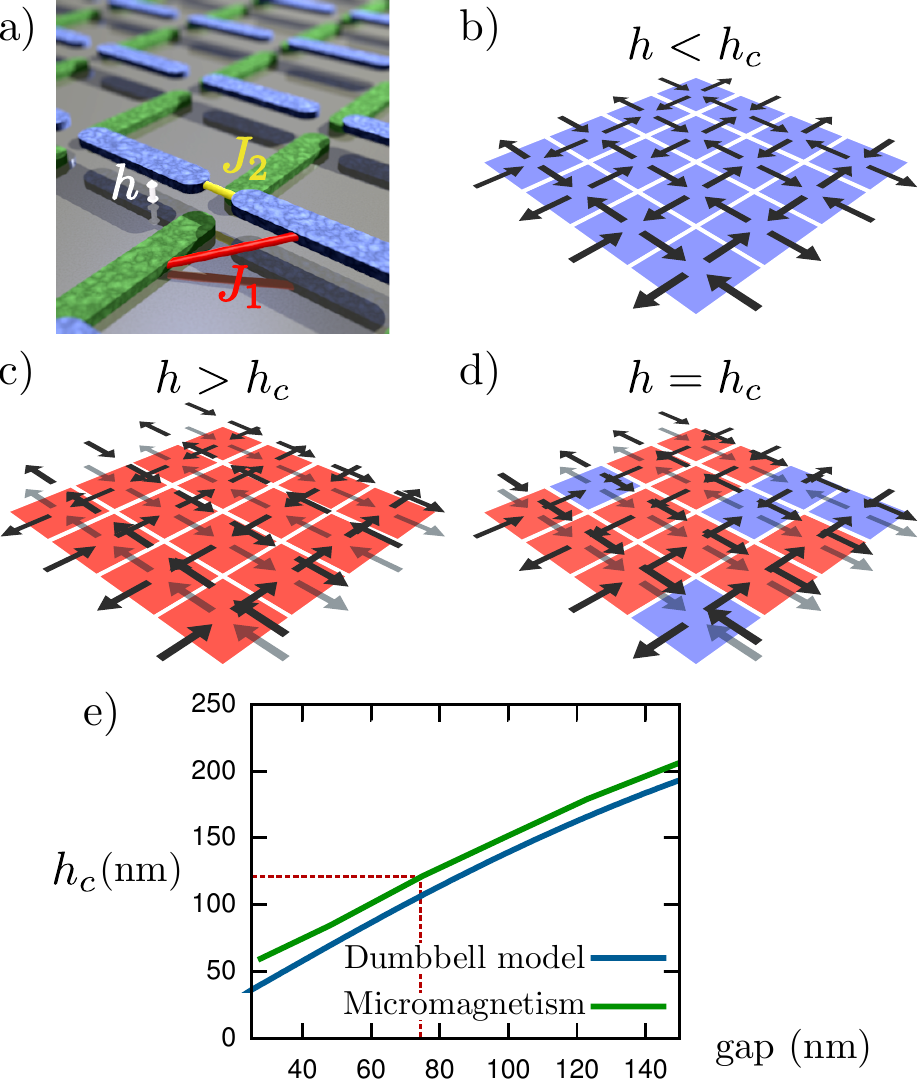}
\caption{\label{shifted_arrays} \textbf{Role of the nearest-neighbor coupling strength.} a) Schematic view of an artificial square ice in which one of the two sub-lattices is shifted vertically by an height offset $h$. The nearest-neighbor coupling strengths between orthogonal ($J_1$) and collinear ($J_2$) nanomagnets are indicated in red and yellow. b-d) Ground states of the models associated with the conditions $J_1 > J_2$ (b), $J_1 < J_2$ (c) and $J_1 = J_2$ (d). e) Plot showing the height offset $h_c$ required to recover an ice-like physics, as a function of the gap separating nearest-neighbor nanomagnets. Results derived from micromagnetic simulations (green curve) and from a dumbbell description (blue curve) of the nanomagnets are compared. The length, width, thickness and gap of our nanomagnets are 500 nm, 100 nm, 30 nm and 75 nm, respectively. More details are provided in the Method section. The red dashed line indicated the $h_c$ value expected in our experiments.}
\end{figure}

Although conceptually simple, square arrays of interacting magnetic nano-islands reveal a very different, and somehow disappointing, behavior than the one predicted by the square ice model. Indeed, in artificial square ice, the nonequivalent nature of the magnetostatic interactions between nearest neighbors [see Fig. \ref{shifted_arrays}a] lifts the energy degeneracy of type-I and type-II vertices. This gives rise to an ordered ground state \cite{Morgan2011, Farhan2013, Porro2013, Kapaklis2014, Drisko2015}, characterized by flux closure magnetic loops with alternating chirality [see Fig. \ref{shifted_arrays}b]. In other words, artificial square ice systems do not mimic the properties of the highly degenerated pyrochlore spin ice materials and cannot be described by the square ice model. Instead, their physics is better described by the Rys F-model \cite{Rys1963} introduced for antiferroelectric crystals and associated to the condition $E_1 < E_2$, where $E_1$ and $E_2$ are the energies of type-I and type-II vertices, respectively. All attempts so far to make an artificial realization of the square ice model were unsuccessful, whether the arrays are field-demagnetized \cite{Wang2006, Nisoli2010, Budrikis2011, Budrikis2012} or thermally annealed \cite{Morgan2011, Farhan2013, Porro2013, Kapaklis2014, Drisko2015}. Recently, more complex topologies involving mixed coordination (such as Shakti lattices) \cite{Gilbert2014} were found to have a high degree of degeneracy and to map on an effective square ice model. However, exotic, many body phenomena such as Coulomb phases and monopole-like excitations on a massively fluctuating uncharged vacuum have not been demonstrated in those realizations. The seminal square ice model still lacks a direct experimental realization that would open the investigation of intriguing collective phenomena present in frustrated magnetic ice systems, via a lab-on-chip approach.

Here, we show that the highly degenerated low-energy manifold expected from the square ice model can be recovered by an appropriate design of the square lattice. The analysis, in real and reciprocal space, of the magnetic configurations resulting from field demagnetization protocols unambiguously reveals that the degeneracy of type-I / type-II vertices is retrieved. In particular, the observed magnetic configurations are disordered and their associated magnetic structure factors show the expected signatures of a spin liquid state with emerging algebraic spin-spin correlations. Besides, we observe local and diluted monopole-like excitations, trapped within an uncharged magnetic background, that can only annihilate with their antiparticules through non trivial spin flips. We thus provide the first artificial realization of the celebrated square ice model, opening new avenues to investigate unconventional magnetic states of matter directly in real space.

\subsection{Tuning the nearest-neighbor coupling strength at will}

To recover the true degeneracy associated with the square ice model, we fabricated a series of artificial square ice systems inspired by the theoretical proposition of M\"{o}ller and Moessner \cite{Moller2006}. The main idea behind that proposition is to reduce the $J_1$ coupling strength, while keeping $J_2$ unchanged, by shifting vertically one of the two sub-lattices of the square array [see Fig. \ref{shifted_arrays}a]. Doing so clearly leaves the $J_2$ coupling unchanged as the distance between the corresponding nanomagnets is not modified. However, the vertical shift between the two sub-lattices increases the distance between perpendicularly-oriented nanomagnets, reducing the $J_1$ coupling strength. Such a height offset $h$ thus offers the possibility to finely tune the $J_{1}/J_{2}$ ratio. If $h=0$, the system is the conventional artificial square ice system, characterized by $J_{1}>J_{2}$ and a magnetically ordered ground state [see Fig. \ref{shifted_arrays}b]. If $h$ is continuously increased, $J_1$ can become infinitively small compared to $J_2$, until one reaches a situation where the horizontal and vertical lines of the square lattice are magnetically decoupled [$J_{1}=0$, see Fig. \ref{shifted_arrays}c]. Therefore, there necessary exists a critical value $h_c$ for which the two coupling coefficients $J_1$ and $J_2$ are equal [see Fig. \ref{shifted_arrays}d]. Based on a dumbbell description of the nanomagnets, M\"{o}ller and Moessner found that $h_c=0.207 \times a$ for $l/a=0.7$, where $l$ is the length of the nanomagnets and $a$ is the lattice parameter. However, this approach neglects key experimental ingredients \cite{Rougemaille2013}: the geometrical properties and the micromagnetic nature of the nanomagnets are not taken into account. Here, we go a step further and determine the critical value $h_c$ from a set of micromagnetic simulations describing the real shape and internal micromagnetic configuration of the nanomagnets used experimentally (see the Method section). The main result of our calculation is that $h_c$ strongly varies with the gap left between neighboring magnetic elements, in qualitative agreement with the estimation deduced from the dumbbell description [see Fig. \ref{shifted_arrays}e], but differs substantially quantitatively.

This result has immediate consequence: to recover the degeneracy of the square ice model, one needs to lithographically pattern arrays of nanomagnets in which the third dimension now plays a key role, bringing artificial spin ice systems from 2 to 3 dimensions. This makes both the fabrication and the imaging of the spin ice architectures much more challenging.

\begin{figure}[H]
\center
\includegraphics[width=16cm]{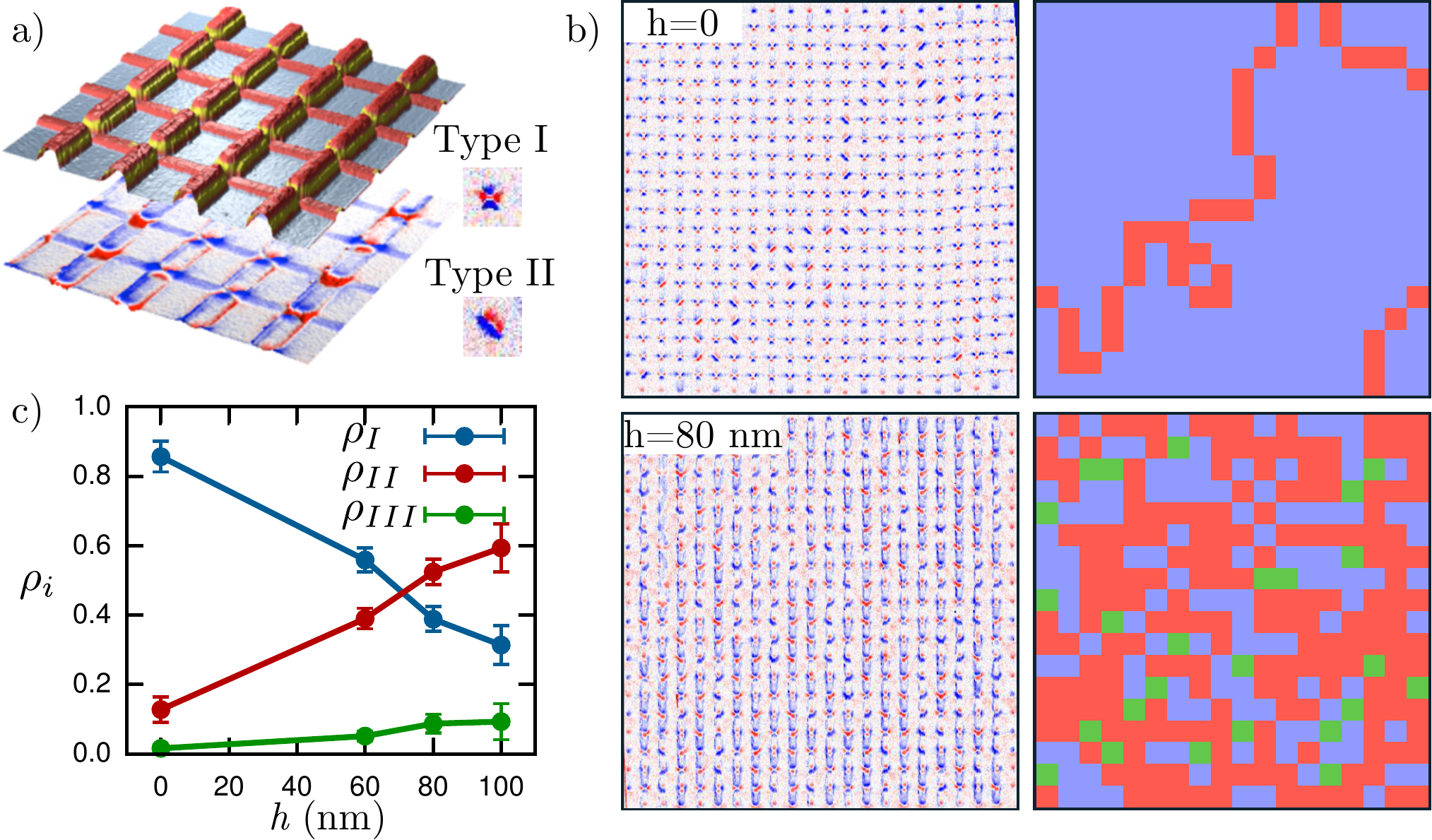}
\caption{\label{experiments} \textbf {Experimental results.} a) Topography (atomic force microscopy) and magnetic (magnetic force microscopy) images of our artificial realization of the square ice model. In the top image, the nanomagnets appear in red, while the bases are colored in yellow and the substrate in grey. In the magnetic image, the magnetic contrast appears in blue and red. Typical contrasts obtained on type-I and type-II vertices in shown in the inset. b) Magnetic images (raw data on the left and their corresponding analysis on the right) for an height offset of $h=0$ nm and $h=80$ nm. In the first case, most of the vertices are of type-I (blue) and a domain boundary separating anti-phase domains is clearly visible (type-II vertices in red). In the second case, the magnetic state appears disordered. c) Analysis of the vertex density $\rho$ as a function of the height offset $h$.}
\end{figure}

Our shifted artificial square ices were fabricated using a two step electron beam lithography process (see the Method section). The first step is dedicated to the design of non-magnetic bases that are used to lift one sub-lattice of nanomagnets. The thickness of the base determines the height offset $h$ finally obtained. The second step consists in depositing the nanomagnets on a square lattice in such a way that one sub-lattice is grown atop the non-magnetic bases, while the other sub-lattice is grown on the substrate. In our samples, the bases are made from gold and titanium, and the nanomagnets result from the deposition of a 30 nm-thick permalloy ($Ni_{80}Fe_{20}$) layer. A 3 nm-thick aluminum capping layer is finally deposited to prevent the nanomagnets from oxidization. The base thicknesses used here are 60, 80 and 100 nm. The nanomagnets have typical dimensions of $500 \times 100  \times 30$ nm$^3$ and the lattice parameter is set to 650 nm [see Fig. \ref{experiments}a]. The distance between the vertex center and the extremity of a nanomagnet (the gap) is thus 75 nm. Our square lattices are composed of 840 nanomagnets. On each sample, a reference square lattice with $h=0$ is patterned for direct comparison with the shifted arrays. Magnetic images were obtained using magnetic force microscopy (MFM) after demagnetizing the arrays in an in-plane oscillating magnetic field with slowly decaying amplitude (see the Method section). All arrays were demagnetized simultaneously to ensure identical field history between samples. The whole demagnetized protocol was repeated several times to improve statistics and to check the reproducibility of the experimental observations.

\subsection{Observation of an algebraic spin liquid state}

\begin{figure}[H]
\center
\includegraphics[width=16cm]{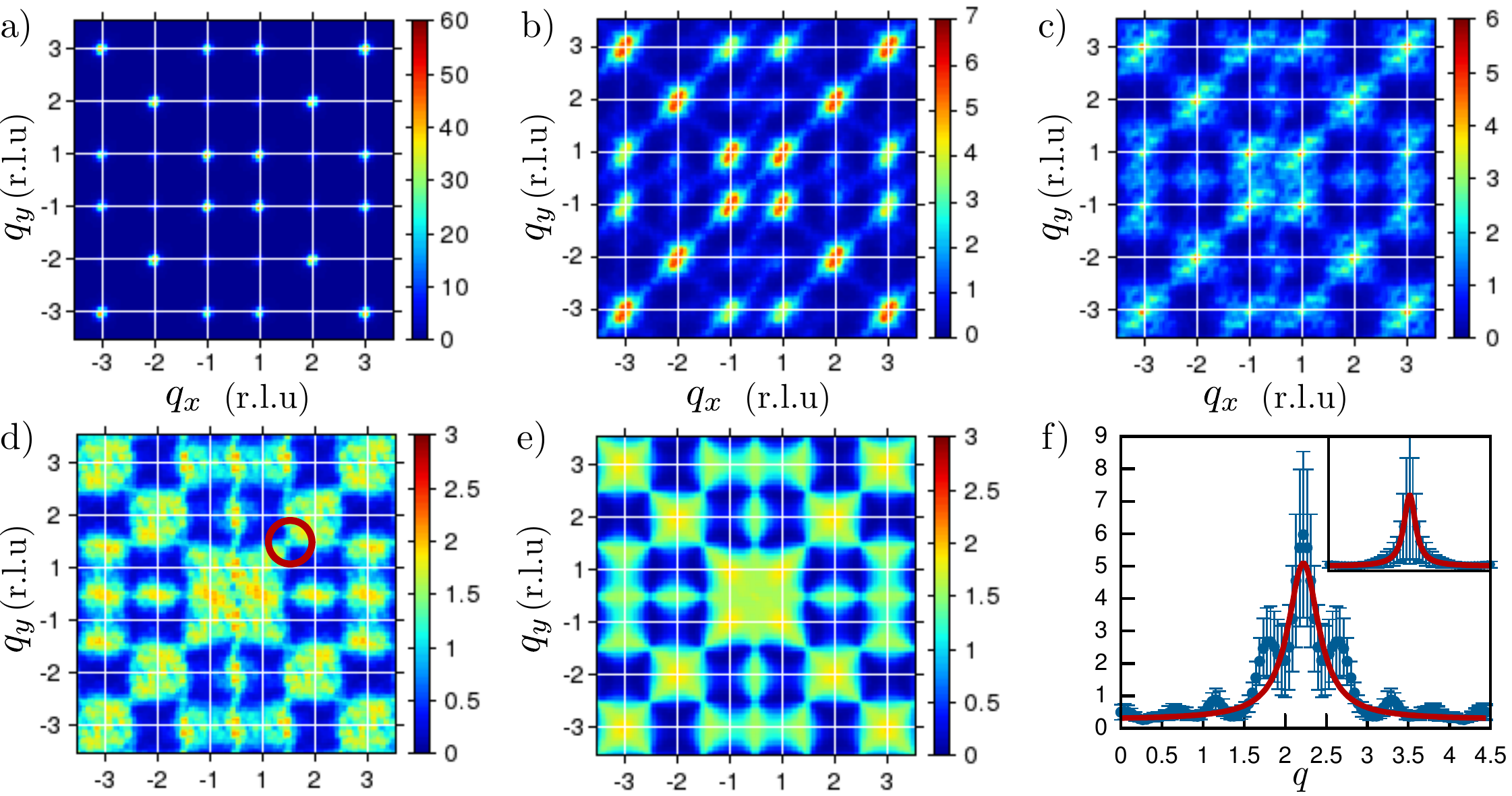}
\caption{\label{structure_factor} \textbf{Magnetic structure factors and pinch point analysis.} a-d) Magnetic structure factors deduced from the experimental images for $h=$ 0, 60, 80 and 100 nm, respectively. e) Computed magnetic structure factor averaged over 1000 random and decorrelated spin configurations satisfying the ice rule (see the Method section). f) Experimental and theoretical (inset) intensity profiles across the pinch point highlighted by a red circle in d).}
\end{figure}

The three shifted arrays studied in this work have been demagnetized four times, and we systematically imaged the reference array ($h=0$) present on each sample to check the efficiency of the field demagnetization protocol. For these 12 realizations, the reference arrays are always found in a magnetic configuration close to the ordered antiferromagnetic (AFM) ground state [Fig. \ref{shifted_arrays}b]. A typical magnetic image is reported in Fig. \ref{experiments}b in which a domain boundary separating two anti-phase domains is observed. Consequently, type-I vertices are present everywhere, except in the domain wall formed by type-II vertices. Our demagnetization protocol is thus efficient and brings the system into a low-energy manifold with large patches of the ground state configuration, similar to what is found in thermally-active artificial spin ices \cite{Morgan2011, Farhan2013, Porro2013, Kapaklis2014, Drisko2015}. On average, the density of type-I, II, III and IV vertices are 86\%, 12.5\%, 1.5\% and 0\%, respectively [see Fig. \ref{experiments}c], and the mean size of type-I domains is about 87 vertices. The residual magnetization is thus low, 3\% typically, in both the vertical and horizontal directions. The computed magnetic structure factor (see the Method section), averaged over the different reference arrays (12 in total), show clear magnetic Bragg peaks located at the corners of the Brillouin zone [see Fig. \ref{structure_factor}a].

Figure \ref{experiments}c also shows the variation of the vertex densities $\rho$ when $h$ is increased and reveals a clear trend: the density of type-I vertices continuously decreases while the density of type-II vertices increases. When $h=60$ nm, the physics is essentially unchanged: type-I vertices dominate and form patches of the AFM ground state, although the average size of the ordered domains drops down to 15 vertices (i.e. is 6 times smaller compared to the reference arrays). The corresponding magnetic structure factor shows a spreading of the magnetic Bragg peaks associated to the AFM ordering, but the intensity remains located at the corners of the Brillouin zone [see Fig. \ref{structure_factor}b]. This result is consistent with the predictions from micromagnetic simulations [see Fig. \ref{shifted_arrays}e], which indicate that the ground state is expected to be the AFM ordered configuration when $h=60$ nm. 

For $h=80$ nm, the population of type-II vertices (52\%) becomes higher than the one of type-I vertices (39\%): type-II patches start to form, while type-I domains have their spatial extension further reduced. The magnetic Bragg peaks in the structure factor have essentially disappeared, meaning that the spin configurations are now highly disordered. More importantly, if the background intensity in the magnetic structure factor becomes more diffuse, it gets structured with geometrical features that strongly resemble those expected from the square ice model [see Figs. \ref{structure_factor}c and \ref{structure_factor}e, and the Method section]. This result is also consistent with the micromagnetic simulations: although the ground state is expected to be the AFM ordered configuration, the magnetic configuration is highly disordered after demagnetizing the array, as $h$ approaches $h_c$ ($J_1$ starts to compare with $J_2$).

The comparison with the square ice model becomes evident for $h=100$ nm [see Figs. \ref{structure_factor}d-e]. Contrary to all works reported so far demonstrating that square lattices of nanomagnets are magnetically ordered in their low-energy manifold, we clearly show here that our artificial square ice is highly disordered. In particular, the magnetic Bragg peaks in the structure factor have now totally disappeared and the diffuse background is strongly structured. We note that the ground state of an artificial square ice with $h=100$ nm is expected to be ordered from the micromagnetic simulations presented above. We might then wonder why ice-like physics is observed for an height offset smaller than predicted. We interpret this difference as a consequence of the kinetic associated to the spin dynamics when the sample is demagnetized under a rotating magnetic field. Indeed, during the demagnetization protocol, spins are reversed through an avalanche process that favors the formation of straight lines (see the Supplementary material). Type-II vertices are then stabilized by the external magnetic field at the expense of type-I vertices, even though type-I vertices have a slightly lower energy. In other words, our protocol shifts the critical value $h_c$ at which the transition to the disordered phase is expected. 

More importantly, our square ice presents all the characteristic of a dipolar algebraic spin liquid, i.e. a correlated disordered magnet within which spin-spin correlations decay like point-dipole interactions \cite{Garanin1999}. In such a spin liquid, there are peculiar locations in the reciprocal space where the magnetic structure factor behaves non analytically and has the shape of a pinch point \cite{Fennell2009}. These pinch points are clearly visible in our experimental map [see circle in Fig. \ref{structure_factor}d] and reveal the fingerprint of a Coulomb phase \cite{Henley2010}. To gain more insights into the observed  physics, it is useful to quantitatively compare, on the experimental and theoretical maps, q-scans of the magnetic structure factor along two orthogonal lines passing through these pinch points (see the Method section). While the theoretical scan reveals a sharp peak associated to a correlation length $\xi$ of the order of a fraction of the lattice size ($\xi_{theo} = 5.2 \pm 4\%$ in lattice parameter unit), the experimental scan displays a broader peak associated to a shorter correlation length ($\xi_{exp} = 4.4 \pm 12\%$ in lattice parameter unit). This points to the presence of local excitations, i.e. to a finite density of classical monopoles within a Coulomb phase.

\subsection{Monopoles trapped within a spin liquid phase}

Experimentally, we observe for all the arrays a similar density of +2 and -2 monopoles (although the arrays do not systematically obey charge neutrality due to their finite size), with a density that increases as we retrieve the degeneracy of the square ice model. We recall here that all the arrays have been demagnetized several times and have felt exactly the same magnetic field history, as they have been always demagnetized all together. The variation of the monopole density we measure is not random and appears to be robust when comparing successive demagnetization protocols. We note that these densities are fairly high when approaching the spin liquid phase [see Fig. \ref{experiments}c], meaning that we do not bring the system into its massively degenerated ground state manifold. Instead, the imaged spin configurations are characteristic of excited states embedded within a Coulomb phase. 

We emphasize that the monopoles we observe in our arrays differ substantially from those that have been visualized so far in artificial square ices \cite{Phatak2011, Farhan2013}. More specifically, all monopoles reported until now are high-energy local configurations evolving in an uncharged, but magnetically ordered vacuum. Two cases may be distinguished from previous literature.

\begin{figure}[H]
\center
\includegraphics[width=15cm]{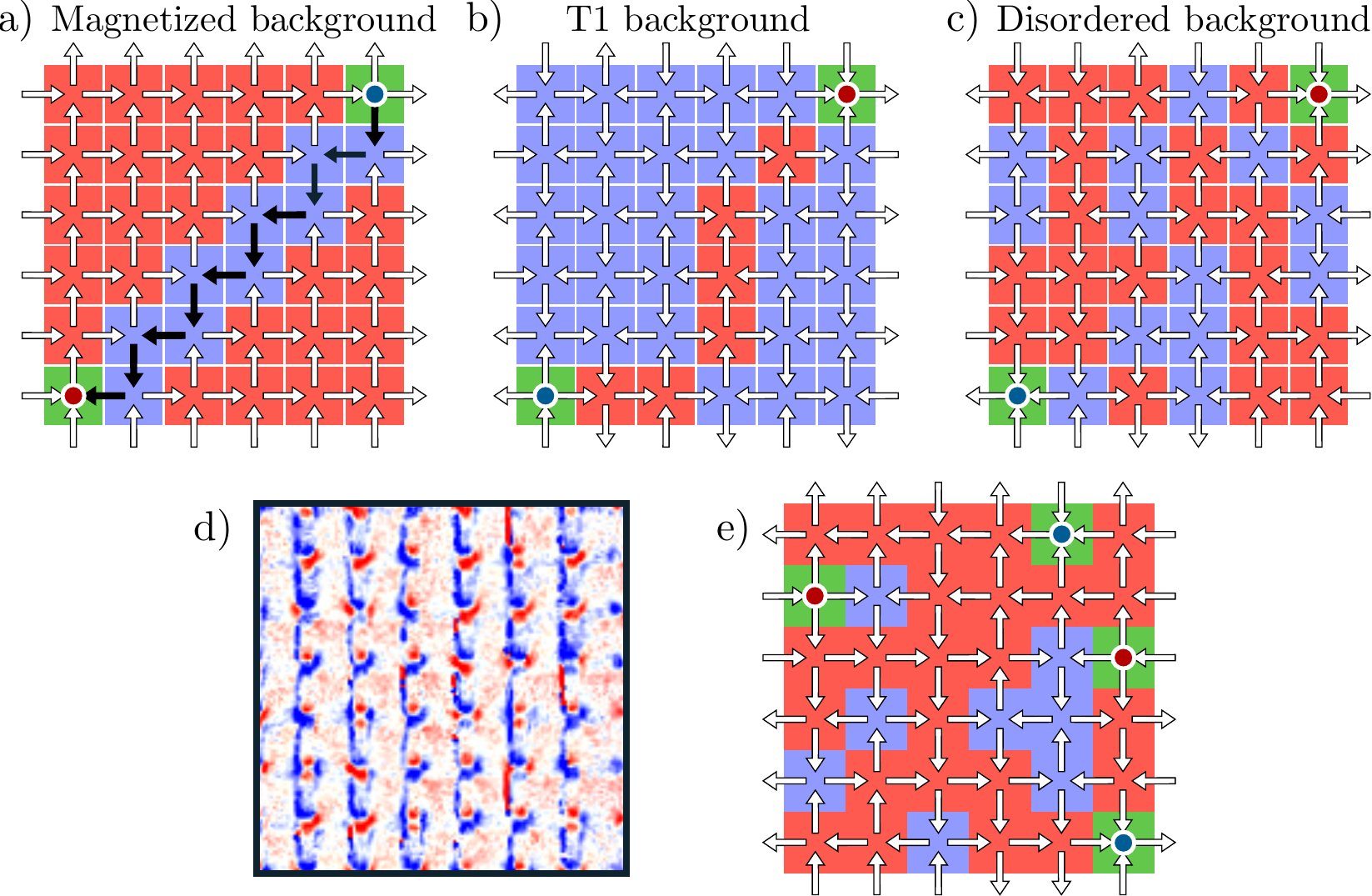}
\caption{\label{monopoles} \textbf{Magnetic monopoles in square ice systems.} Monopole / anti-monopole pair in a magnetically saturated background (a), in the AFM ground state (b) and within a disordered manifold (c). (d) Experimental spin configurations in the case $h=100$ nm showing two pairs of oppositely charged monopoles. (e) Analysis of the configuration reported in d). Monopoles appear as red and blue circles.}
\end{figure}

1) The first observation of monopoles in artificial square ice was achieved after saturating the arrays using a magnetic field applied along a (11)-like direction and by subsequently applying a field in the opposite direction with an amplitude close to the coercive field of the systems. The protocol then induces random nucleations of monopoles and triggers an avalanche process \cite{Phatak2011}. This protocol leads to a unidirectional motion of the monopoles that leave behind them chains of reversed spins often referred to as Dirac strings. Similar results have been obtained in thermally active arrays that have been magnetically saturated \cite{Farhan2013}. There, monopoles are metastable objects, created on purpose, embedded within a magnetically saturated state, i.e. a spin configuration containing mainly type-II vertices [see Fig. \ref{monopoles}a].

2) The second observation of monopoles in artificial square ice was done in arrays approaching the AFM ordered ground state after being demagnetized or annealed \cite{Farhan2013}. Monopoles do not necessarily move along straight lines, but are always confined within a domain boundary separating anti-phase ground state domains made of type-I vertices [see Fig.\ref{monopoles}b]. The monopoles are not free particules but topological defects allowing AFM domains to grow, i.e. they are charged objects embedded within a magnetically ordered state.

In both cases, monopoles evolve within an ordered magnetic configuration characterized by a magnetic structure factor that only contains magnetic Bragg peaks. Our system has a very distinct behavior: the monopoles we observe are free to move into a spin liquid state, i.e. a massively degenerated, disordered low-energy manifold [see Fig. \ref{monopoles}c]. They are thus particule-like objects present in a diffuse but structured magnetic structure factor, free of any Bragg peak [see Fig. \ref{structure_factor}d]. An example of experimental configuration is reported in Fig. \ref{monopoles}d and schematized in Fig. \ref{monopoles}e. Two pairs of oppositely charged monopoles are present in this disordered magnetic configuration containing type-I and type-II vertices. We want to stress that it is not possible to determine the path these monopoles have followed during the demagnetization process, as the trace of reversed spins (often referred to as Dirac string) has been somehow erased by the magnetic disorder. This is in sharp contrast with the two cases mentioned above where monopoles evolve within an ordered spin configuration and in which the influence of the field or temperature can be unambiguously visualized [see Figs. \ref{monopoles}a-b]. Here, there is no mean to know what was the trajectory  of the magnetic monopoles. Because of this incapability to track their motion, it is in fact not even possible to pair two oppositely charged monopoles.

This result raises interesting, crucial questions. If similar artificial shifted square ice systems could be made thermally active, we could envision to investigate in real space and in real time the dynamics of these deconfined, interacting quasi-particules. We can then wonder if a typical distance between oppositely charged monopoles would established at thermodynamic equilibrium and if this distance could be linked to the correlation length deduced from the analysis of the pinch point width. We could also imagine to study how these quasi-particules nucleate, propagate and annihilate with their anti-particule, and directly observe how their interactions affect the disordered background.

\subsection{Summary and prospects}

We showed in this work that our shifted magnetic square lattices offer the possibility to tune at will the six-vertex model from its ordered (Rys F-model) regime into its disordered phase, and in particular to realize the seminal square ice model for the first time. At the peculiar point where $J_1=J_2$, the demagnetized samples we studied exhibit the signatures of an excited Coulomb phase, i.e. a magnetic phase characterized by a finite density of magnetic monopoles embedded within an algebraic spin liquid. The fabrication of thermally-active shifted square ice would certainly be the next step as it would provide with the possibility to investigate the thermodynamics as well as the dynamics of these exotic low-energy manifolds and the recombination of their topological excitations.

Finally, we would like to point out that our work demonstrates that artificial loop models, such as the square ice model, are not beyond reach, thanks to lithography engineering. These models have numerous extensions in very different fields of research, such as polymer physics, topological quantum computing, self-avoiding random walks or Schramm-Loewner evolution. Implementing loop models is therefore of very broad interest in physics and chemistry, and our contribution illustrates that magnetic versions of these loop models are now accessible experimentally.

\subsection{Methods}

\subsubsection{Micromagnetic simulations}
Our micromagnetic simulations are based on a finite difference approach, i.e. the system is discretized into rectangular cells. The 3D solver of the OOMMF free software from NIST has been used in this work \cite{OOMMF}. We computed the four energy levels $E_i$ ($i\in 1,2,3,4$) of an isolated square vertex composed of four permalloy nanomagnets in magnetostatic interaction. The nanomagnets have dimensions of 500 $\times$ 100 $\times$  30 nm$^3$. The gap $g$ between the nanomagnets is defined as the distance between the nanomagnet extremities and the center of the vertex. In all calculations, the exchange stiffness is set to 10 pJ/m, the magnetocrystalline anisotropy is neglected, spontaneous magnetization $M_S$ is $8 \times 10^5$ (1.0 T) and the damping coefficient is set to 1 to speed up convergence. Considering that the volume of a nanomagnet is $1.44 \times 10^{-21}$ m$^3$, our nanomagnets carry a magnetic moment about $\mu=11.52 \times 10^{-16}$ A.m$^2$.  To limit the influence of numerical roughness, the mesh size has been reduced to 3 $\times$ 3 $\times$ 15 nm$^3$. No qualitative difference has been observed when reducing the mesh size in the $z$ direction.

\subsubsection{Dumbbell model}
\begin{figure}[H]
\center
\includegraphics[width=9cm]{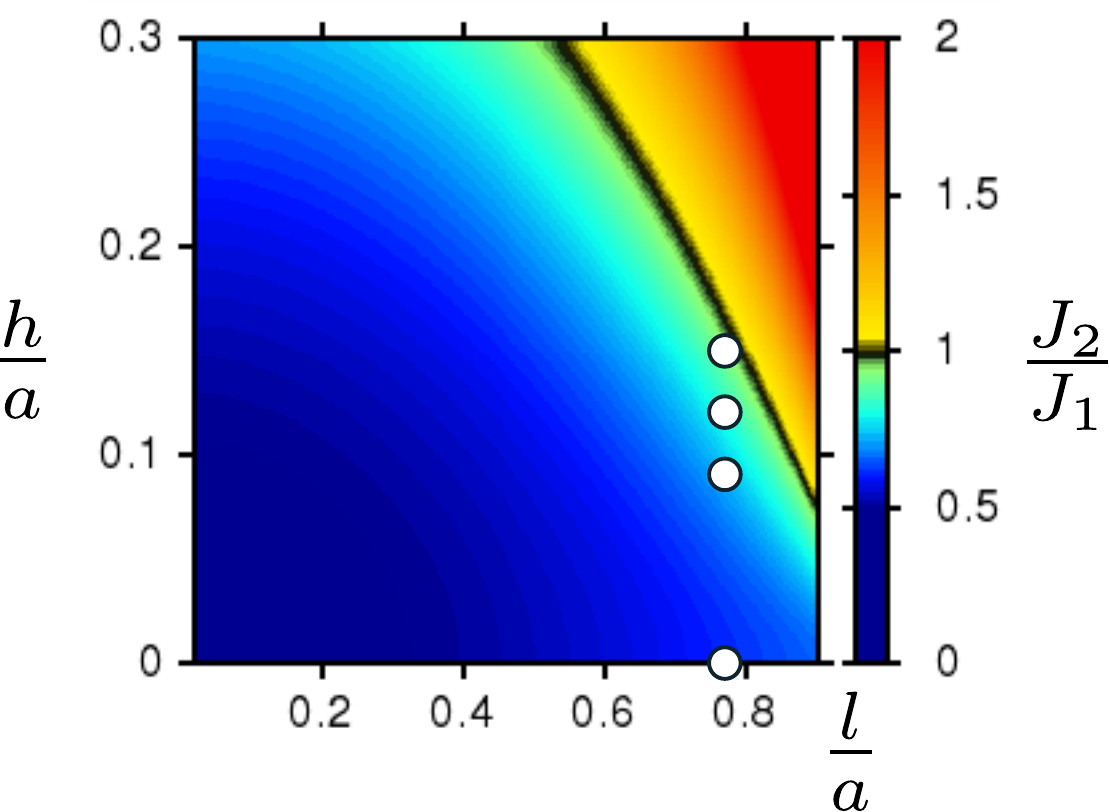}
\caption{\label{map} \textbf{Dumbbell description of the nanomagnets.} Map of the ${J_1}/{J_2}$ ratio as a function of $l$ and $h$ for an isolated vertex. The condition $J_1=J_2$ is verified on the dark line. The white dots indicate the values corresponding to the different samples studied in this work.}
\end{figure}

In Figure \ref{shifted_arrays}e, we plot the value of the critical height $h_c$ as a function of the gap between neighboring nanomagnets in two different cases: within a full micromagnetic approach or using a dumbbell description. For the dumbbell description, we followed the same procedure as the one reported by M\"oller and Moessner \cite{Moller2006}. Their results have been reproduced and are represented in Figure \ref{map}, showing the ratio ${J_1}/{J_2}$ as a function of $l$ and $h$, where $l$ is the length of the nanomagnets, i.e. the distance between the two magnetic charges. The condition $J_1=J_2$ is verified on the dark line.

\subsubsection{Sample fabrication}
\begin{figure}[H]
\center
\includegraphics[width=10cm]{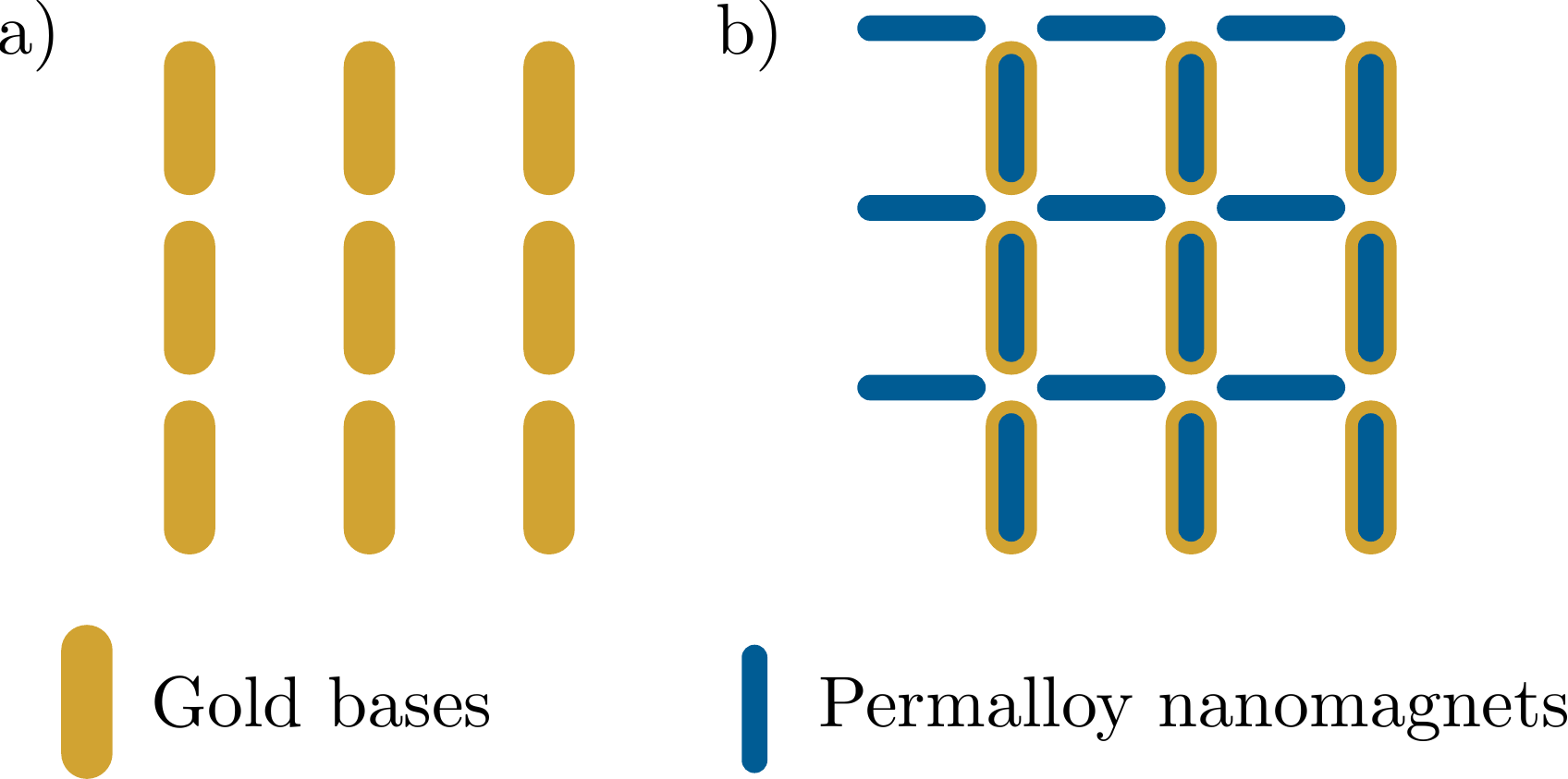}
\caption{\label{fabrication} \textbf{Illustration of the two step e-beam lithography process.}}
\end{figure}

The shifted artificial square spin ices were fabricated using a two step e-beam lithography process [see Fig. \ref{fabrication}]. The first step is dedicated to the design of the non-magnetic titanium/gold bases. Their shape has been optimized to maximize the probability to obtain a successful alignment of the nanomagnets deposited in the second step. After exposing and developing a PMMA [poly(methyl methacrylate)] layer, the bases were deposited by e-beam evaporation. To obtain a strong contrast in the scanning electron microscope through the PMMA resist, the top of the bases was made of a 50 nm-thick gold layer. The Ti thickness is then adjusted to obtain the desired height offset between the two sub-lattices. An ultrasound assisted lift-off at $80^{\circ}$ C in the remover reveals the bases.
A new layer of PMMA resist is then spin-coated on the sample. Arrays of nanomagnets were patterned atop the base areas and deposited by e-beam evaporation after resist developing. The ferromagnetic layer is made of a 30 nm-tick $Ni_{80}Fe_{20}$ and a 3 nm-tick Aluminum capping layer was deposited to avoid oxidization. A 5 nm-thick Ti layer between the bases and the permalloy layer is used to enhance its adherence. Finally, a similar lift-off process removes the unwanted material of the samples.
For the two steps, a 150 nm-thick PMMA layer was spin-coated on a (100) silicon substrate. E-beam lithography steps were done using a Raith LEO SEM. The two layers were aligned manually in translation and rotation using adapted cross marks. Our success rate using this method was around 20$\%$ for the nanomagnet sizes employed here. Ti, Al, Au and $Ni_{80}Fe_{20}$ were deposited with an evaporation rate of 0.1 nm/s (pressure $10^{-5}$ mbar).

\subsubsection{Magnetic imaging}
Magnetic images were obtained using a NT-MDT magnetic force Microscope. Homemade low-moment magnetic tips were used to avoid magnetization reversal in the nanomagnets while scanning the arrays. The magnetic layer of the tips was made from a 30 or 50 nm-thick CoCr alloy. Prior to their imaging, the samples were demagnetized using an in-plane oscillating field (250 mHz sine function) with decaying amplitude, while the sample is put in rotation at a typical frequency of several tens of Hz. A very slow field ramp was used to decrease the amplitude of applied external magnetic field from 100 mT (well above the coercive field of our nanomagnets) to zero in 72 hours. As demonstrated by the magnetic configurations obtained on the different reference samples ($h=0$ nm), our protocol efficiently brings the arrays within their low-energy manifold.

\subsubsection{Magnetic structure factor}
We define the magnetic structure factor like it is done in neutron scattering experiments, where the spin correlations perpendicular to the diffusion vector $\vec{q}$ are measured. We thus define a perpendicular spin component $\vec{S_{i\alpha}}^\perp$:

\begin{equation}
\vec{S_{i\alpha}^\perp}=\vec{S_{i\alpha}}-(\vec{\hat{q}}.\vec{S_{i\alpha}})\vec{\hat{q}}
\label{sip}
\end{equation}

where $\vec{\hat{q}}$ is the unit vector along the diffusion vector $\vec{q}$ :
\[\vec{\hat{q}}= \frac{\vec{q}}{||\vec{q}||} \]

\begin{figure}[h]
\includegraphics[width=\textwidth]{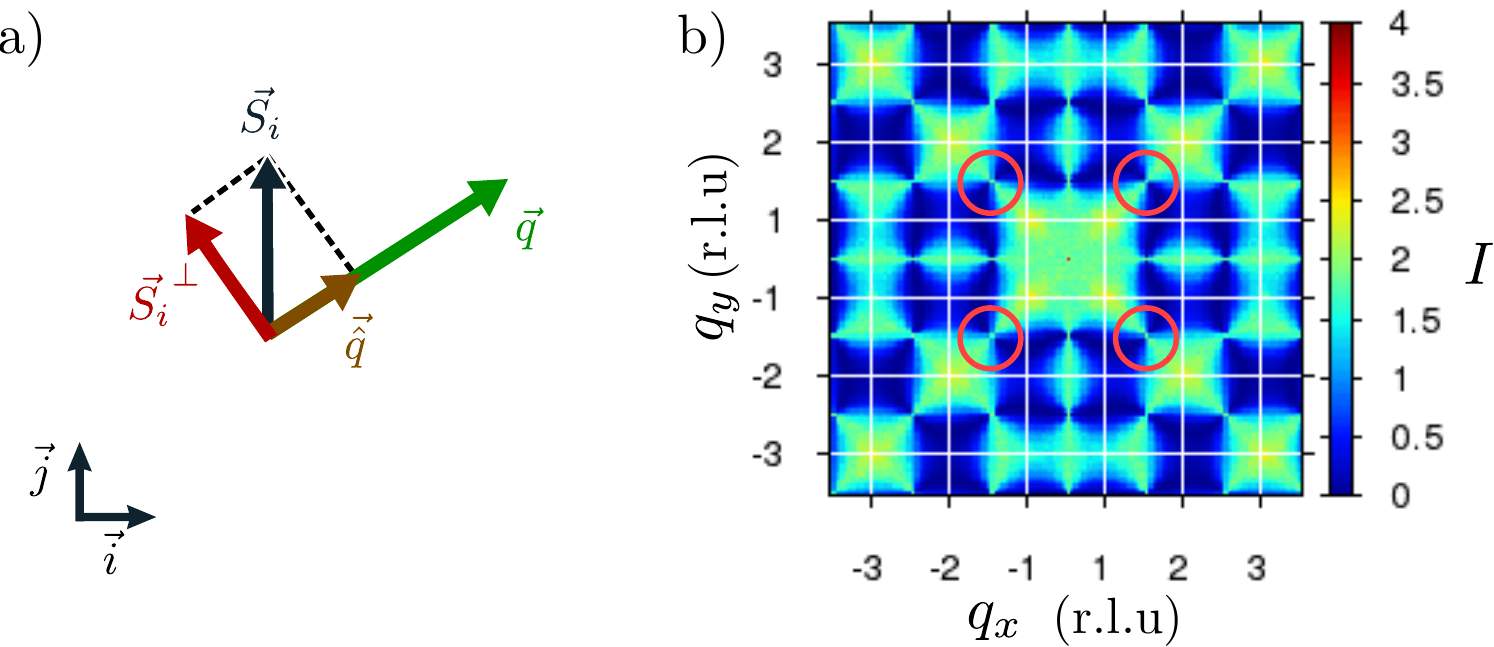}
\caption{\label{Structure_factor_2} \textbf{Magnetic structure factor of the square ice model.} (a) Sketch of the vectors involved in Eq. (\ref{sip}). (b) Magnetic structure factor for an ideal Lieb model, computed for 1000 low-energy states made of $N=840$ spins. Red circles indicate the region of interest for the intensity profiles reported in figure \ref{pp}.}
\end{figure}

Figure \ref{Structure_factor_2}a shows the geometrical construction of the vectors involved in Eq. \ref{sip}. We note $I(\vec{q})$ the intensity scattered at the location $\vec{q}$ in the reciprocal space:

\begin{equation}
I(\vec{q})=\frac{1}{N} \sum_{i,j=1}^{\frac{1}{2}N} \sum_{\alpha,\beta=1}^2 \vec{S_{i\alpha}^\perp}.\vec{S_{j\beta}^\perp} \exp{i\vec{q}.\vec{r}_{i\alpha, j\beta}}
\label{I}
\end{equation}

In this expression $i$ and $j$ scan all the $\frac{1}{2}N$ unity cells, and $\alpha, \beta$ the two sites of each cell. The following discussion aims at obtaining a more convenient form for (\ref{I}), allowing a direct calculation starting from a magnetic configuration. One can show that $I(\vec{q})$ can be split into two parts, $I^{\slash{}\slash{}}$ and $I^\perp$, giving $I=I^{\slash{}\slash{}}-I^\perp$ with:

\begin{align}
I^{\slash{}\slash{}}(\vec{q}) &= \frac{1}{N} \sum_{\alpha,\beta=1}^2 g_\alpha(\vec{q})g_\alpha^*(\vec{q}) \\
I^\perp(\vec{q}) &= \frac{1}{N} \sum_{\alpha,\beta=1}^2 p_\alpha(\vec{q})p_\beta^*(\vec{q}) \\
\text{where~~~~~} g_\alpha(\vec{q}) &= \sum_{i=1}^N \sigma_{i\alpha} \exp{i\vec{q}.\vec{r}_{ij}} \\
p_\alpha(\vec{q}) &= \sum_{i=1}^N (\vec{\hat{q}}.\vec{S_{i\alpha}}) \exp{i\vec{q}.\vec{r}_{ij}}
\end{align}

where $\sigma_{i\alpha}$ is the Ising variable ($\pm 1$) of the site ($i\alpha$). Under that split form, $I$ is a real quantity. To compute $I(\vec{q})$ diagrams in reciprocal space, we calculate the quantity $I= g_1^2+g_2^2 -(p_1 + p_2)^2$ at several $\vec{q}$ locations. The magnetic structure factor reported in Figure \ref{Structure_factor_2}b is composed of a matrix of $120 \times 120$ points covering an area $q_x,q_y \in [-6\pi:6\pi]$. This area is 36 times larger than the first Brillouin zone.

\subsection{Generating low-energy magnetic configurations}

\begin{figure}[H]
\center
\includegraphics[width=0.5\textwidth]{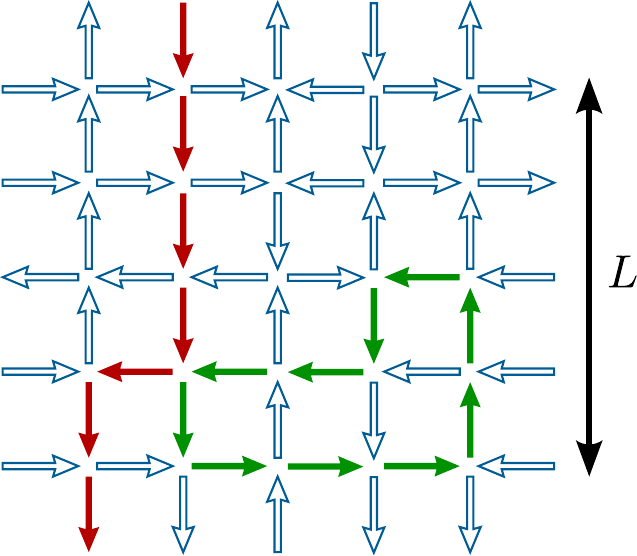}
\caption{\label{loop} \textbf{Loop flips in the square lattice.} Schematics illustrating the open (red arrows) and close (green arrows) spin loops used to generate a low-energy configuration representative of the massively degenerated ground state manifold of the (Lieb) square ice model. The lattice contains 840 spins and the number $N$ of loops that are flipped between two decorrelated configurations is set to 840.}
\end{figure}

The magnetic structure factor of the (Lieb) square ice model was computed by averaging 1000 low-energy spin configurations satisfying the ice rule everywhere. To do so, we start from a magnetically saturated configuration and then flip a number $N$ of spin loops chosen randomly. These loop flips are necessary to ensure that all the generated spin configurations satisfy the ice rule. Because our lattice has free boundary conditions, the procedure leads to open (crossing the array) or close loops [see Fig. \ref{loop}]. Both loops are used to generate a low-energy spin configuration. To decorrelate the initial (saturated) and final spin configurations, we take $N$ of the order of the number of spins present in the array (840).

\subsection{Pinch points and correlation length}

\begin{figure}[H]
\center
\includegraphics[width=0.8\textwidth]{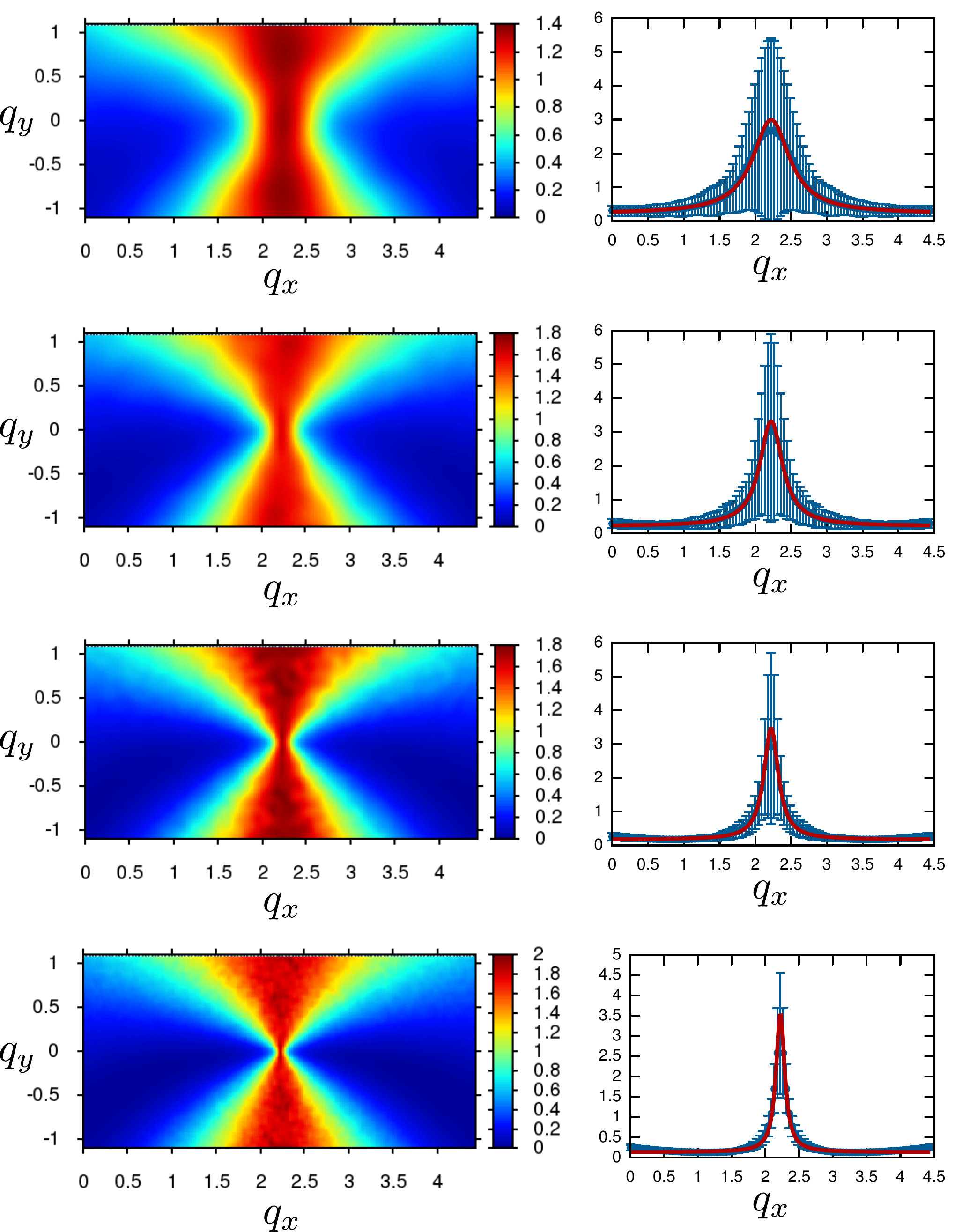}
\caption{\label{pp} \textbf{Analysis of the pinch points.} Maps of the pinch points indicated by a red circle in Fig. \ref{Structure_factor_2}b and associated intensity profiles along $q_y=0$, for different lattice sizes. From top to bottom : $L$=10, 20, 40, 80. The coordinates $(q_x,q_y)$ are relative to the intensity profile and do not correspond to the real axes of the reciprocal space. The red curves represent the Lorentzian fits derived from Eq.\ref{fit}.}
\end{figure}

The magnetic structure factor shown in Figure \ref{Structure_factor_2}b is averaged over 1000 ice rule configurations. Pinch points located at the center $\Gamma$ of the Brillouin zone are clearly visible, indicating the existence of a Coulomb phase and algebraic spin-spin correlations. The finite size of our arrays has consequences on this magnetic structure factor, in particular on the width of the pinch points. Figure \ref{pp} shows the influence of the lattice size $L$ on the width of the pinch points. As expected, the pinch point get narrower as the lattice size is increased [see intensity profiles in Figure \ref{pp}]. This width of the pinch points can be linked to a correlation length $\xi$ in the system \cite{Fenell2009}. This correlation length $\xi$ can be extracted from a Lorentzian fit of the intensity profile passing through a pinch point:
\begin{equation}
I(q)=A \dfrac{\xi^{-2}}{(q-q_0)^2+ \xi^{-2}} +B
\label{fit}
\end{equation}

In Eq.\ref{fit}, $A$ and $B$ are constants, $q_0$ is associated to the location of the pinch point in reciprocal space, and $q$ is the diffusion vector. The correlation lengths deduced for different lattice sizes are reported in table \ref{tabcor}. Error bars represents the variability within the 1000 sampled spin states.

\begin{table}[H]
\centering
	\begin{tabular}{l|c|c|c|c}
		$L$ & 10 & 20 & 40 & 80 \\
		\hline
		$\xi$ & $3.2 \pm 2\%$ & $5.2 \pm 4\%$ & $8.2 \pm 7\%$ & $13.0 \pm 13\%$ \\
	\end{tabular}	
	\caption{\label{tabcor} Correlation lengths extracted from the intensity profiles.}
\end{table}

\newpage

\bigskip

\pagebreak

\subsection{Acknowledgements}

This work was supported by the Agence Nationale de la Recherche through project through projects no. ANR12-BS04-009 'Frustrated'. The authors also acknowledge support from the Nanofab team at the Institut NEEL and warmly thank S. Le-Denmat and O. Fruchart for technical help during AFM/MFM measurements.

\subsection{Author contributions}
B.C. and N.R. conceived the project. Y. P. was in charge of the sample preparation, the magnetic imaging measurements and the analysis of the data. All authors contributed to the preparation of the manuscript.

\subsection{Additional information}
Supplementary Information accompanies this paper.

Competing financial interests: The authors declare no competing financial interests.

\end{document}